\documentclass[a4paper,11pt]{article}

\usepackage{jcappub}

\usepackage{slashed}
\usepackage{youngtab}

\newcommand\fverb{\setbox\fverbbox=\hbox\bgroup\verb}
\newcommand\fverbdo{\egroup\medskip\noindent%
            \fbox{\unhbox\fverbbox}\ }
\newcommand\fverbit{\egroup\item[\fbox{\unhbox\fverbbox}]}
\newbox\fverbbox
\newcommand{\nablaslash}{\not{\hbox{\kern-3pt $\nabla$}}}

\title{Revisiting slow-roll inflation in nonminimal derivative coupling with potentials}

\author{Yun~Soo~Myung$^{a}$,}
\author{Taeyoon~Moon$^{a}$}
\affiliation{$^{a}$Institute of Basic Science and Department of
Computer
Simulation, Inje University,\\
Gimhae 621-749, Korea}
\author{and~Bum-Hoon Lee$^{b}$}
\affiliation{$^{b}$Department of Physics and Center for Quantum Spacetime, Sogang University,\\
Seoul 121-742, Korea}

\emailAdd{ysmyung@inje.ac.kr} \emailAdd{tymoon@inje.ac.kr}
\emailAdd{bhl@sogang.ac.kr}

\abstract{We investigate the  slow-roll inflation in the nonminimal
derivative coupling (NDC) model with exponential, quadric, and
quartic potentials. It was known that this model provides an
enhanced slow-roll inflation induced by gravitationally enhanced
friction even for a steep exponential potential. In the phase
portrait, the inflationary attractor is described by the slow-roll
equation. Introducing the autonomous form, the inflation is regarded
as an emergence from the saddle point  and it leaves this fixed
point along the slow-roll equation. We show explicitly that if one
uses the NDC with potentials,  the slow-roll inflation is easier to
be implemented than the canonical coupling with the same
potentials.}

\begin{document}

\maketitle \flushbottom
\section{Introduction}

The nonminimal derivative coupling
(NDC)~\cite{Amendola:1993uh,Sushkov:2009hk}  was achieved by
coupling the inflaton kinetic term to the Einstein tensor such that
the friction is enhanced gravitationally at higher
energies~\cite{Germani:2010gm}. This gravitationally enhanced
friction mechanism
 is a powerful tool to increase friction of an inflaton rolling down its own potential without introducing
 new  degrees of freedom  unlike hybrid inflation. Actually, this NDC  makes a steep (non-flat) potential adequate for inflation without
possessing  higher-time derivative terms (ghost
state)~\cite{Germani:2011ua,Germani:2011mx}. It is well-known that
the exponential potential provides a power-law inflation and thus,
it cannot account for an inflationary theory since the inflation
never ends and an additional mechanism is required to stop
it~\cite{PU}. The kinetic coupling flattens
 the potential effectively as well as it  increases friction.
 Recently, this coupling allowed  inflation to take place
for a wide range of values $\lambda$ in the (non-flat) exponential
potential $V=V_0e^{-\lambda \phi}$ which implies that it features a
natural exit from inflation~\cite{Dalianis:2014nwa}.

In this paper, we study the autonomous dynamical system of an
homogeneous and isotropic configuration of a scalar field non
minimally coupled to gravity, as in
\cite{Copeland:1997et,Leon:2014rra,Leon:2008de,Fadragas:2014mra,
Cai:2014uka,Felder:2002jk,Kofman:2002cj,Hindmarsh:2011hx,Contillo:2011ag}.
We found that in the two dimensional dynamical system of the field
value and its derivative, the space of initial conditions providing
inflationary attractors is larger in NDC than in the canonical case.
Specifically we have investigated NDC models with exponential,
quadric, and quartic potentials since non-flat potentials of
exponential and quartic potentials with canonical coupling
(CC)~\cite{Yang:2015pga} are in tension with the Planck
data~\cite{Ade:2015lrj}. Especially, the exponential potential is
regarded as a testing potential because it could  provide a
power-law (eternal) inflation in the CC model while it could provide
a slow-roll inflation in the NDC model~\cite{Tsujikawa:2012mk}.

\section{An inflation model of NDC}

Let us consider an inflation model whose action includes NDC of
scalar field $\phi$ with a potential
term~\cite{Germani:2010gm,Germani:2011ua,Germani:2011mx,Feng:2014tka}
\begin{eqnarray} \label{mact}
S_{\rm}=\frac{1}{2}\int d^4x \sqrt{-g}\Big[M_{\rm
P}^2R+\frac{1}{\tilde{M}^2}G_{\mu\nu}\partial^{\mu}\phi\partial^{\nu}\phi-2V(\phi)\Big],
\end{eqnarray}
where $M_{\rm P}$ is a reduced Planck mass, $\tilde{M}$ is a mass
parameter, and  $G_{\mu\nu}=R_{\mu\nu}-g_{\mu\nu}R/2$ is the
Einstein tensor. Here, we do not include a canonical coupling (CC)
term like as a conventional combination of
$(g_{\mu\nu}-G_{\mu\nu}/\tilde{M}^2)\partial^\mu\phi\partial^\nu
\phi$~\cite{Tsujikawa:2012mk,Skugoreva:2013ooa} because this
combination won't make the analysis transparent (see Appendix A for
the autonomous system for CC+NDC).

Varying the action (\ref{mact}) with respect to  the metric tensor
$g_{\mu\nu}$ leads to  the Einstein equation
\begin{equation} \label{einseq}
G_{\mu\nu} =\frac{1}{M_{\rm P}^2} T_{\mu\nu}^{\rm NDC},
\end{equation}
where $T_{\mu\nu}^{\rm NDC}$ takes the complicated  form
\begin{eqnarray}
 T_{\mu\nu}^{\rm
NDC}&=&\frac{1}{\tilde{M}^2}\Big[\frac{1}{2}R\nabla_{\mu}\phi\nabla_{\nu}\phi
-2\nabla_{\rho}\phi\nabla_{(\mu}\phi R_{\nu)}^{\rho}
+\frac{1}{2}G_{\mu\nu}(\nabla\phi)^2-R_{\mu\rho\nu\sigma}\nabla^{\rho}\phi\nabla^{\sigma}\phi
\nonumber
\\&&\hspace*{5em}-\nabla_{\mu}\nabla^{\rho}\phi\nabla_{\nu}\nabla_{\rho}\phi
+(\nabla_{\mu}\nabla_{\nu}\phi)\nabla^2\phi\nonumber\\
&&
\label{em1}-g_{\mu\nu}\Big(-R^{\rho\sigma}\nabla_{\rho}\phi\nabla_{\sigma}\phi+\frac{1}{2}(\nabla^2\phi)^2
-\frac{1}{2}(\nabla^{\rho}\nabla^{\sigma}\phi)\nabla_{\rho}\nabla_{\sigma}\phi
\Big)\Big].
\end{eqnarray}
Here, we note that even though fourth-order derivative terms are
present in $T^{\rm NDC}_{\mu\nu}$, there is no ghost state which
means that any higher-time derivative term more than two is not
generated\footnote{We check easily  that the two terms for
$\ddot{\phi}^2$ in the second line of Eq.(\ref{em1}) cancel against
each other and the last two terms in the last line of Eq.(\ref{em1})
do too.}. On the other hand, the scalar equation is derived to be
\begin{equation} \label{scalar-eq}
-\frac{1}{\tilde{M}^2}G^{\mu\nu}\nabla_{\mu}\nabla_{\nu}\phi-V'=0,
\end{equation}
where the prime (${}^{\prime}$) denotes derivative with respect to
$\phi$.

 In this work, we consider a spatially flat spacetime  by introducing cosmic time $t$ as
\begin{eqnarray} \label{deds1}
ds^2~=~\bar{g}_{\mu\nu}dx^\mu dx^\nu~=~-dt^2+a^2(t)\delta_{ij}dx^idx^j,
\label{deds2}
\end{eqnarray}
where $a(t)$ is a scale factor. In this spacetime, two Friedmann
equations  and scalar equation become
\begin{eqnarray}
H^2&=&\frac{1}{3M_{\rm P}^2}\Big[\frac{9H^2}{2\tilde{M}^2}\dot{\phi}^2+V\Big],\label{Heq}\\
&&\nonumber\\
\dot{H}&=&-\frac{1}{2M_{\rm
P}^2}\Big[\dot{\phi}^2\Big(\frac{3H^2}{\tilde{M}^2}-\frac{\dot{H}}{\tilde{M}^2}\Big)-\frac{2H}{\tilde{M}^2}\dot{\phi}\ddot{\phi}
\Big],\label{dHeq}\\
&&\nonumber\\
&&\hspace*{-4em}\frac{3H^2}{\tilde{M}^2}\ddot{\phi}+3H\Big(\frac{3H^2}{\tilde{M}^2}+\frac{2\dot{H}}{\tilde{M}^2}\Big)\dot{\phi}+V'=0\label{seq},
\end{eqnarray}
where $H=\dot{a}/a$ is the Hubble parameter and  the overdot
($\dot{}$) denotes derivative with respect to time $t$. We observe
from  (\ref{Heq}) that the energy density for the NDC is  positive
(ghost-free) \cite{Germani:2011bc}.

Now we define two slow-roll
parameters as
\begin{eqnarray}
\epsilon_N=-\frac{\dot{H}}{H^2},~~~~~\delta_N=\frac{\ddot{\phi}}{H\dot{\phi}}.\label{scon}
\end{eqnarray}
Imposing  slow-roll conditions ($\epsilon_{N}\ll1,~\delta_{N}\ll1$),
Eqs. (\ref{Heq})-(\ref{seq})  can be written approximately as
\begin{eqnarray}
H^2&\simeq&\frac{1}{3M_{\rm P}^2}V,
\label{Heqs1}\\
\dot{H}&\simeq&-\frac{3H^2}{2M_{\rm
P}^2\tilde{M}^2}\dot{\phi}^2,\label{dHeqs1}\\
\hspace*{-5em}3H\dot{\phi}&\simeq&-\frac{\tilde{M}^2}{3H^2}V'. \label{seqs1}
\end{eqnarray}
We note that the slow-roll parameter $\epsilon_N$ can be written  as
\begin{eqnarray}
\epsilon_N~\simeq~\epsilon_N^{V}~\simeq~\epsilon_N^{H},
\end{eqnarray}
where the potential $\epsilon_N^{V}$ and  Hubble slow-roll
parameters $\epsilon_N^{H}$ are given by
\cite{Germani:2011ua,Germani:2010ux}
\begin{eqnarray}\label{slpu}
\epsilon_N^{V}=\frac{M_{\rm
P}^2\tilde{M}^2}{6}\frac{V'^2}{V^2H^2},~~~~~\epsilon_N^{H}=\frac{3}{2M_{\rm
P}^2\tilde{M}^2}\dot{\phi}^2.
\end{eqnarray}
The end of inflation can be identified with the end of slow-roll
regime. Therefore, we may  define the value of field at the end of
inflation $\phi_f$ by making use of $\epsilon^V_N=1$.

\section{Dynamical analysis}
In this section, we wish to perform the dynamical analysis for
(\ref{Heq})-(\ref{seq}). For this purpose,  one introduces with some
dimensionless quantities and  obtains  their autonomous system.
Before we proceed, it is instructive to investigate the  canonical
coupling (CC) model for comparison and exercise.

\subsection{Analysis of CC  models}

In this case, the corresponding field equations are given by
\begin{eqnarray}
H^2&=&\frac{1}{3M_{\rm P}^2}\Big[\frac{1}{2}\dot{\phi}^2+V\Big],\label{Heqc}\\
\dot{H}&=&-\frac{1}{2M_{\rm
P}^2}\dot{\phi}^2,\label{dHeqc}\\
&&\hspace*{-2em}\ddot{\phi}+3H\dot{\phi}+V'=0\label{seqc}.
\end{eqnarray}
It is well-known that in  the slow-roll approximation, they reduce
to
\begin{eqnarray}
H^2&\simeq&\frac{V}{3M_{\rm P}^2},\label{sHeqc}\\
\dot{H}&=&-\frac{1}{2M_{\rm
P}^2}\dot{\phi}^2,\label{sdHeqc}\\
3H\dot{\phi}&\simeq& -V'\label{sseqc},
\end{eqnarray}
which are obtained by imposing  slow-roll condition of
$\epsilon_C=-\dot{H}/H^2\ll1$ and
$\delta_C=\ddot{\phi}/H\dot{\phi}\ll1$. Here we have $
\epsilon_C\simeq \epsilon_C^V\simeq \epsilon_C^H$ with
\begin{eqnarray}\label{slpc}
\epsilon_C^{V}\equiv\frac{M_{\rm P}^2}{2}\frac{V'^2}{V^2},~~~~~\epsilon_C^{H}\equiv\frac{\dot{\phi}^2}{2M_{\rm
P}^2H^2}.
\end{eqnarray}
We note that the NDC parameters of $\epsilon^{V,H}_N$
 (\ref{slpu}) are slightly different from the  CC parameters
$\epsilon^{V,H}_C$ (\ref{slpc}). In particular,
$\epsilon^{V}_N=\epsilon^{V}_C\times\tilde{M}^2/(3H^2)$ implies that
when $\tilde{M}^2/(3H^2)\ll1$, inflation in the NDC can happen more
easily than the CC case.

 To perform the dynamical analysis for the cosmological evolution
 equations
(\ref{Heqc})-(\ref{seqc}), we consider the dimensionless variables
\begin{eqnarray}
x\equiv\frac{\dot{\phi}}{M_{\rm P}\sqrt{6} H},
~~~y\equiv\frac{\sqrt{V}}{M_{\rm P}\sqrt{3}
H},~~~\alpha_{V}\equiv\sqrt{6}M_{\rm
P}\frac{V'}{V},~~~\Gamma\equiv\frac{VV''}{V'^2}.\label{dimless}
\end{eqnarray}
The connections between $(x,\alpha_V)$  and
($\epsilon^{H}_C,\epsilon^V_C$)  are given by
\begin{equation}\label{xalpha}
x^2=\frac{\epsilon_C^H}{3},~~\alpha^2_V=12 \epsilon_C^V.
\end{equation}
Then, the first Friedmann equation (\ref{Heqc}) becomes a constraint
equation
\begin{eqnarray}
 x^2+y^2=1,
\end{eqnarray}
which implies that the  variable  $y$ can be eliminated from the
dynamical equations.

 It is found that (\ref{Heqc})-(\ref{seqc})
 lead to the
autonomous form for ${\bf X}=(x,\alpha_V)$
\begin{eqnarray}
\frac{dx}{dN}&=&3(x^2-1)\Big(x+\frac{\alpha_{V}}{6}\Big),\label{delx}\\
\frac{d\alpha_{V}}{dN}&=&\alpha_{V}^2x(\Gamma-1)\label{delalphav}
\end{eqnarray}
with $N=\ln a$.  Fixed points of first-order autonomous system  are
given by $dx/dN=0$ and $d\alpha_V/dN=0$.  By a fixed point we mean
that $x$ and $\alpha_V$ do not  change as the universe evolves.
Perturbing these points leads to stable point (attractor), saddle
point, and unstable point (repeller). Then, one can classify these
according to their eigenvalue solution to perturbed equations: all
negative (stable point); all positive (unstable), different signs
(saddle point).

At this stage, it is worth noting  that the slow-roll trajectory can
be found  by considering  two different ways:  One is found by
solving  the slow-roll equations (\ref{sHeqc})-(\ref{sseqc}) and the
other is found by solving  the autonomous system (\ref{delx}) and
(\ref{delalphav}) in the slow-roll approximation. The former
indicates an inflationary attractor in phase portrait
$(\phi,\dot{\phi})$, while the latter shows that inflation is
regarded as  ``an emergence from the saddle point and it can leave
this fixed point along the slow-roll equation". For this purpose, we
introduce three potentials: exponential, quadric (chaotic), and
quartic potentials. The first one is regarded as a testing potential
because it cannot provide  a slow-roll inflation in the CC model,
whereas it could provide a slow-roll inflation in the NDC model.

We should distinguish between inflationary attractor in
($\phi,\dot{\phi}$) and attractor of stable fixed point  in
($x,\alpha_V$). To this end, we note that  slow-roll equations
(\ref{sHeqc})-(\ref{sseqc}) are combined to give
\begin{eqnarray}\label{dotphi}
\dot{\phi}\simeq-M_{\rm P}^2\frac{HV'}{V},
\end{eqnarray}
which describes an inflationary attractor.   This  can be expressed
in terms of $x$ and $\alpha_V$ defined in (\ref{dimless}) as
\begin{eqnarray}\label{xa}
x\simeq-\frac{\alpha_V}{6}.
\end{eqnarray}
We note that Eq.(\ref{xa}) satisfies
$\epsilon_{C}^H\simeq\epsilon_{C}^{V}$ which yields
$x^2\simeq\alpha_V^2/36$, when substituting (\ref{xalpha}) into
$\epsilon_{C}^{H,V}$. Also, (\ref{xa}) describes  a slow-roll line
solution
 to an approximate equation
\begin{eqnarray}
\frac{dx}{dN}&\simeq&x+\frac{\alpha_{V}}{6}\simeq 0,\label{delxs}
\end{eqnarray}
which was  found  from  (\ref{delx}) by  taking into account the
slow-roll condition of $\epsilon_C^{H}=3x^2\ll1$. This implies that
any fixed point can be  defined  only for $\Gamma=1$
($\alpha_V$=const) or $x=0$ ($\alpha_V$=0), while the slow-roll line
is defined without imposing $\alpha_V$=const. Here, a slow-roll line
is  given by a function [$\alpha_V(x)=-6x$] which exists during
inflation. Hence, we identify the inflationary attractor in phase
portrait ($\phi,\dot{\phi}$) with the slow-roll line (equation) in
stream flow ($x,\alpha_V$). The slow-roll approximation reduces the
order of the system equations by one and thus, its general solution
contains one less initial condition. It works only because  the
solution to the full equations has  an attractor property
eliminating the dependence on the extra parameter.
\begin{figure}[t!]
\begin{center}
\begin{tabular}{c}
\includegraphics[width=.90\linewidth,origin=tl]{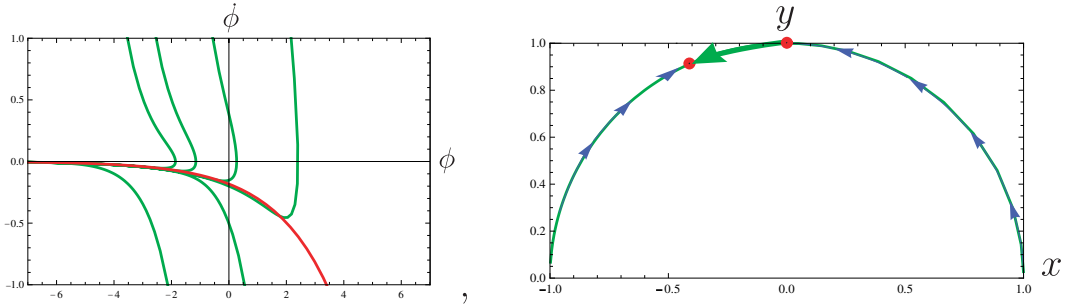}
\end{tabular}
\end{center}
\caption{The phase portrait $(\phi,\dot\phi)$ [left] and stream flow
$(x,y)$ [right] for $V=V_0e^{\lambda\phi}$ with $V_0=0.1$,
$\lambda=1$ and  $M_{\rm P}=1$.   Left panel shows the trajectories
(green curves) with various initial conditions  toward the
inflationary attractor (red curve). This red curve is mapped into a
fixed point of red dot at $(-0.408,0.912)$ in the right panel. This
corresponds to a stable fixed point (attractor), while the point
(0,1) denotes a saddle point. The green curves in the left are
mapped on the green curves on the upper circle in the right.}
\end{figure}

For complete analysis, we choose an explicit potential. \\

$(i)~V=V_0e^{\lambda\phi}$\\
\\
For an exponential potential, the inflaton velocity (\ref{dotphi})
is given by
\begin{eqnarray}\label{pdpp}
\dot{\phi}\simeq-M_{\rm
P}\lambda\sqrt{\frac{V_0}{3}}e^{\frac{\lambda}{2}\phi}.
\end{eqnarray}
Solving  (\ref{Heqc})-(\ref{seqc}) numerically for various initial
conditions, they show  the trajectories. Also, the slow-roll
equation  (\ref{pdpp}) may indicate an inflationary trajectory [see
Fig.1 (left)]. However, this case corresponds to
$\epsilon_C^V=M^2_{\rm P}/2$ (power-law inflation), implying that it
cannot be a complete inflationary model because inflation never ends
with the potential $V=V_0e^{\lambda \phi}$.
 In
order to see it more clearly, we have to know what happens in the
stream plot where fixed points are included naturally. In this case,
we have $\Gamma=1(\alpha_V$=const). Therefore, we could not make a
stream flow on $(x,\alpha_V)$.  Instead, we have the stream flow on
($x,y$).  We find a corresponding fixed point of
$(x,y)=(-\alpha_{V}/6,\sqrt{1-\alpha_{V}^2/36})=(-0.408,0.912)$ in
Fig.1 (right) which turns out to be an attractor (stable fixed
point),  in addition to the (0,1)-saddle point. Clearly, it
indicates that the potential $V=V_0e^{\lambda\phi}$ in the CC model
cannot provide a complete slow-roll inflation because power-law
inflation never ends. This means that there is no spiral sink  which
indicates the end of inflation  in the left of Fig. 1.
\\
\\
$(ii)~V=V_0\phi^2$\\
\\
In the case of chaotic potential, the inflationary attractor
(\ref{dotphi}) is given by
\begin{eqnarray}\label{solph11}
\dot{\phi}\simeq\left\{\begin{array}{ll}
-2M_{\rm P}\sqrt{\frac{V_0}{3}},
~~(\phi>0) \label{soo1}\\
+2M_{\rm P}\sqrt{\frac{V_0}{3}},~~(\phi<0) \label{soo2}\end{array}\right.,
\end{eqnarray}
which corresponds to a positive (negative) constant $\dot\phi$-line
for $\phi<0$ ($\phi>0$) in the ($\phi,\dot{\phi}$) picture.
\begin{figure}[t!]
\begin{center}
\begin{tabular}{c}
\includegraphics[width=.80\linewidth,origin=tl]{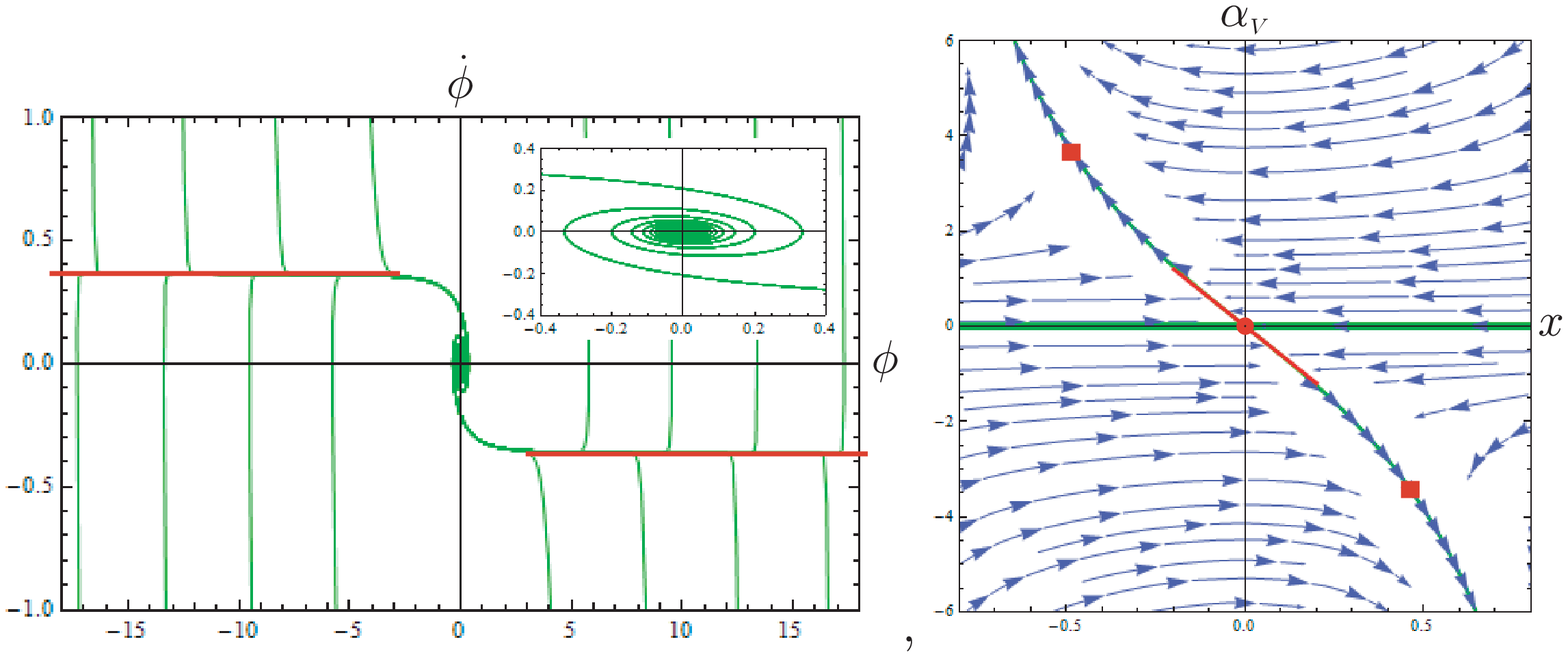}
\end{tabular}
\end{center}
\caption{The phase portrait $(\phi,\dot\phi)$ [left] and stream flow
$(x,\alpha_V)$ [right] for  $V=V_0\phi^2$ with $V_0=0.1$ and $M_{\rm
P}=1$. Left panel shows the trajectories toward  the inflationary
attractor (red lines) for $\phi\lessgtr0$ including a spiral sink.
These red lines correspond to a red line of $\alpha_{V}=-6x$ in the
right panel. Also, the green flows in the right  show the numerical
plot for ($x,\alpha_{V}$) when using (\ref{Heqc})-(\ref{seqc}).
These flows indicate that the inflation is emergent from the saddle
point ($\bullet$) (0,0) and is realized along the red line. Finally,
the inflation ends at the point ({\footnotesize$\blacksquare$})
($\mp0.46, \pm3.48$), corresponding to $\epsilon_{C}^{V}=1$.}
\end{figure}

Figure 2 shows a typical picture based on numerical computation.
When the universe evolves according to (\ref{Heqc})-(\ref{seqc}) and
(\ref{delx})-(\ref{delalphav}), the inflationary attractor  and
slow-roll line are given by (\ref{solph11}) in the phase portrait
($\phi,\dot{\phi}$) [Fig.2 (left)
panel]~\cite{Felder:2002jk,Kofman:2002cj,PU} and  (\ref{xa}) in
stream flow ($x,\alpha_{V}$) [Fig.2 (right) panel], respectively.
There are three phases in the CC case~\cite{Donoghue:2007ze}:
initially, kinetic energy dominates and due to the rapid decrease of
the kinetic energy the trajectory runs quickly to the inflationary
attractor line (\ref{solph11}). All initial trajectories are
attracted to this line, which is the key feature of slow-roll
inflation. Finally, at the end of inflation, there is inflaton decay
and reheating (spiral sink).

On the other hand, the inflation can be realized as an emergence
from the potential-dominated fixed point $(0,0)$ which is a saddle
point because orbits near it are attracted along one direction and
repelled along another direction. The stream flows in the right
 of Fig. 2 can leave this fixed point only along the red line
of $\alpha_V=-6x$ corresponding to the slow-roll line (\ref{xa}) and
the inflation ends at the point ($\mp0.46, \pm3.48$), which
corresponds to $\epsilon_{C}^{V}=1$ with $\alpha_V=\pm3.48$. We note
here that $x$-axis is the stable manifold of the saddle point, while
$\alpha_V=-6x$ is the unstable manifold of the saddle point.
\\
\\
$(iii)~V=V_0\phi^4$\\
\\
The inflationary attractor (\ref{dotphi}) for this case is given by
\begin{eqnarray}\label{solph3}
\dot{\phi}\simeq
-4M_{\rm P}\sqrt{\frac{V_0}{3}}\phi.
\end{eqnarray}
\begin{figure}[t!]
\begin{center}
\begin{tabular}{c}
\includegraphics[width=.80\linewidth,origin=tl]{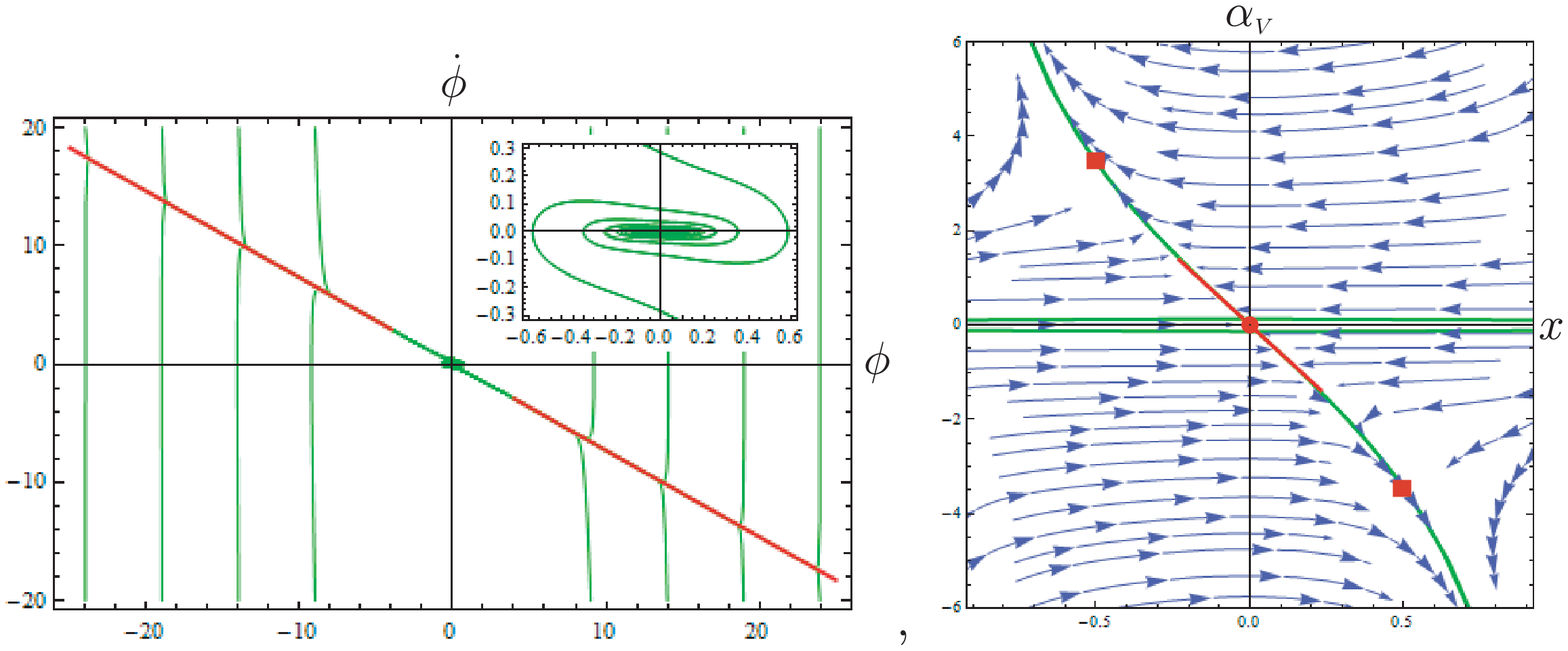}
\end{tabular}
\end{center}
\caption{The phase portrait $(\phi,\dot\phi)$ [left] and stream flow
$(x,\alpha_V)$ [right] for  $V=V_0\phi^4$ with $V_0=0.1$, $M_{\rm
P}=1$. Left panel shows the trajectories toward  the inflationary
attractor (red lines) followed by spiral sink. These lines
correspond to a red line of $\alpha_{V}=-6x$ in the right panel.
Also, the green flows in the right  show  the numerical plot for
($x,\alpha_{V}$) when using (\ref{Heqc})-(\ref{seqc}). These flows
indicate that the inflation is emergent from the saddle point
($\bullet$) (0,0) and is realized along the red line. Finally, the
inflation ends at the point ({\footnotesize$\blacksquare$})
($\mp0.50, \pm3.48$), which corresponds to $\epsilon_{C}^{V}=1$.}
\end{figure}
Figure 3 shows that when solving  (\ref{Heqc})-(\ref{seqc}) and
(\ref{delx})-(\ref{delalphav}) numerically, the inflationary
attractor (decreasing function) is  given by the slow-roll equation
(\ref{solph3}) followed by a spiral sink in the phase portraits
($\phi,\dot{\phi}$) [Fig.3 (left)] and the red line (\ref{xa})  in
stream flows ($x,\alpha_{V}$) [Fig.3 (right)], respectively.
Inflation can be realized as an emergence from the saddle point
$(0,0)$. The inflation  can leave this fixed point only along the
red line of $\alpha_V=-6x$ corresponding to the slow-roll line
(\ref{xa}) and it ends at the point ($\mp0.50, \pm3.48$), which
corresponds to $\epsilon_{C}^{V}=1$ with $\alpha_V=\pm3.48$.

\subsection{Analysis of NDC models}

Now we turn to the NDC case. To perform the dynamical analysis for
(\ref{Heq})-(\ref{seq}),
 we  consider the  dimensionless parameters $u$ and $\alpha_u$ in addition to $y$ and $\Gamma$ defined in (\ref{dimless})
\begin{eqnarray}
u\equiv\sqrt{\frac{3}{2}}\frac{\dot{\phi}}{M_{\rm P}\tilde{M}},
~~~\alpha_{u}=\sqrt{6}\frac{M_{\rm
P}\tilde{M}V'}{VH}.\label{dimless1}
\end{eqnarray}
The connections between $(u,\alpha_u)$  and
($\epsilon^{H}_N,\epsilon^V_N$)  are given by
\begin{equation}\label{ualpha}
u^2=\epsilon_N^H,~~\alpha^2_u=36 \epsilon_N^V.
\end{equation}
Then, the constraint equation (\ref{Heq}) becomes
\begin{eqnarray}
 u^2+y^2=1,
\end{eqnarray}
which implies that one may eliminate  $y$ from the dynamical
equations. It turns out that  (\ref{Heq})-(\ref{seq})
 lead to the autonomous form for ${\bf X}=(u,\alpha_u)$
\begin{eqnarray}
\frac{du}{dN}&=&(2\epsilon_{N}-3)u-\frac{1}{2}\alpha_{u}(1-u^2),\label{delu}\\
\frac{d\alpha_{u}}{dN}&=&\epsilon_{N}\alpha_{u}+\frac{1}{3}(\Gamma-1)u\alpha_{u}^2,\label{delalphau}
\end{eqnarray}
where $\epsilon_{N}=-\dot{H}/H^2$ satisfies the relation
\begin{eqnarray}
(1+u^2)\epsilon_{N}=3u^2+\frac{1}{3}\alpha_{u}u(1-u^2).
\end{eqnarray}
On the other hand, we note that slow-roll equations
(\ref{Heqs1})-(\ref{seqs1}) give us the inflaton velocity
\begin{eqnarray}\label{dotphiu}
\dot{\phi}\simeq-M_{\rm P}^2\tilde{M}^2\frac{V'}{3HV},
\end{eqnarray}
which corresponds to
\begin{eqnarray}\label{ua}
u\simeq-\frac{\alpha_u}{6}.
\end{eqnarray}
We note that Eq.(\ref{ua}) satisfies
$\epsilon_{N}^H\simeq\epsilon_{N}^{V}$ which yields
$u^2\simeq\alpha_u^2/36$, when substituting (\ref{ualpha}) into
$\epsilon_{N}^{H,V}$. Also, it is worth to mention  that
Eq.(\ref{ua}) corresponds to Eq.(\ref{xa}) in CC. Furthermore,
Eq.(\ref{ua}) can be realized as a slow-roll line solution to
(\ref{delu}) when implementing slow-roll condition of
$\epsilon_N^{H}=u^2\ll1$  which takes the form
\begin{eqnarray}
\frac{du}{dN}&\simeq&-3u-\frac{\alpha_{u}}{2}\simeq 0.\label{delxus}
\end{eqnarray}
Differing with the CC model,  there exists an upper limit of
$\dot{\phi}^2$
\begin{eqnarray} \label{cond-NDC}
0 <\dot{\phi}^2 < \phi_{c}^2\equiv\frac{2}{3}M_{\rm P}^2\tilde{M}^2,
\end{eqnarray}
which comes from Eq.(\ref{Heq}) yielding
$H^2(1-\dot{\phi}^2/\phi^2_c)=V/3M^2_{\rm P}$ where the left-handed
side should be positive for a positive potential. This  may be
regarded as another representation of slow-roll condition
($\epsilon^H_N =u^2\ll 1$).
\begin{figure}[t!]
\begin{center}
\begin{tabular}{cc}
\includegraphics[width=.45\linewidth,origin=tl]{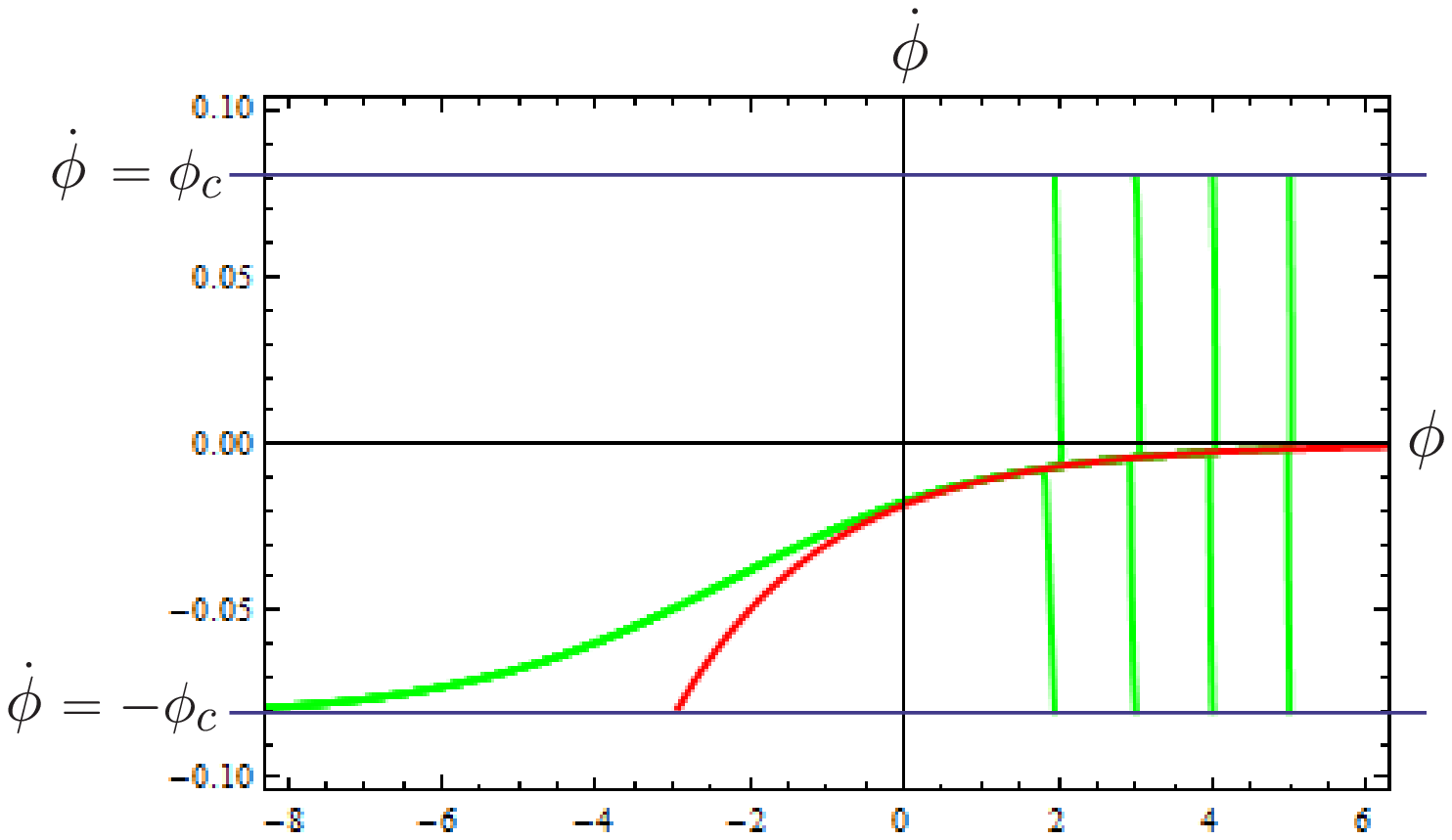}
, &\includegraphics[width=.40\linewidth,origin=tl]{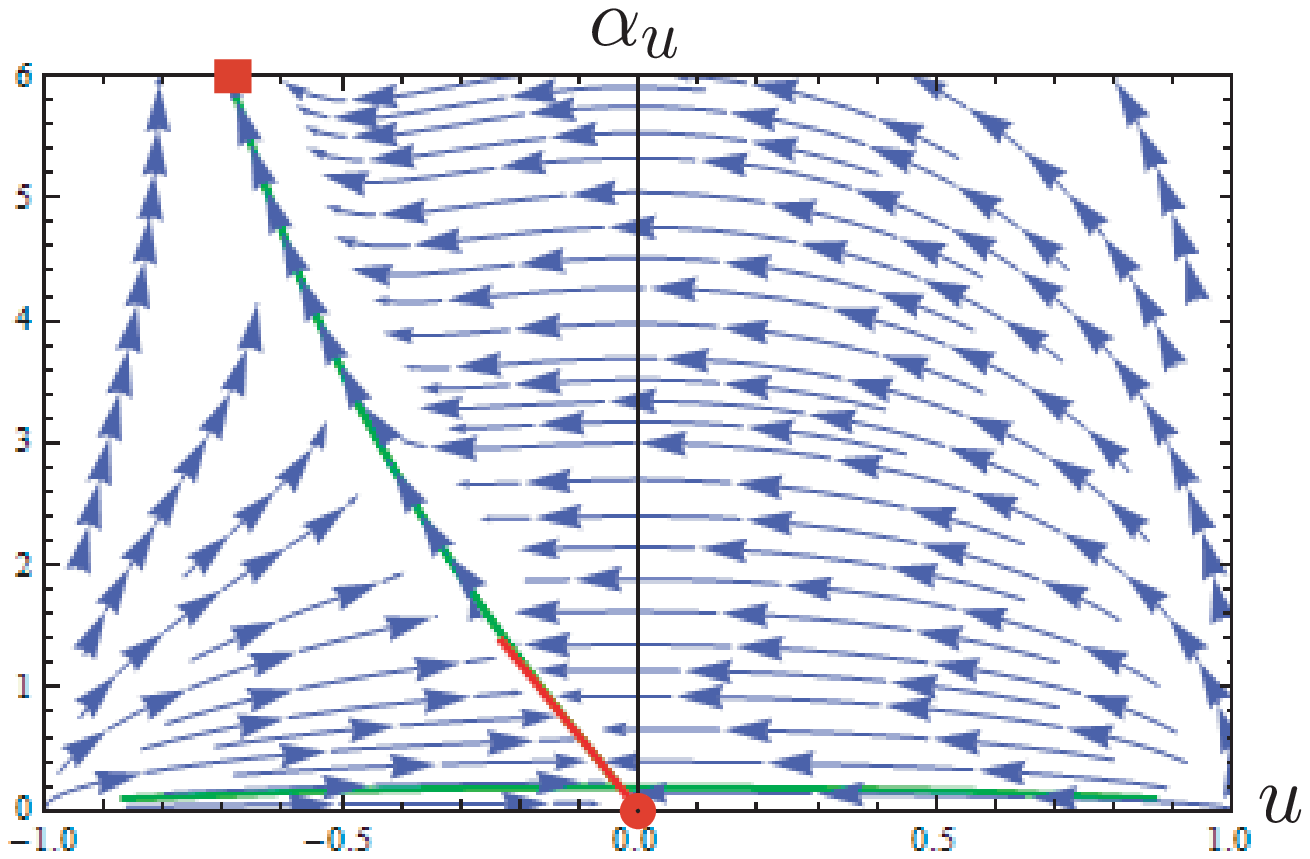}
\end{tabular}
\end{center}
\caption{The phase portrait $(\phi,\dot\phi)$ [left] and stream flow
$(u,\alpha_u)$ [right] for $V=V_0e^{\lambda\phi}$ with $V_0=0.1$,
$\lambda=1$, $M_{\rm P}=1$, $\tilde{M}=0.1$, and $\phi_c=M_{\rm
P}\tilde{M}\sqrt{2}/{\sqrt{3}}\simeq0.08$. Left panel shows the
trajectories toward the inflationary attractor (red curve). Here is
no stable limited cycle. This curve corresponds to a red line of
$\alpha_{u}=-6u$ in the right panel.  Also, the green flows in the
right  show the numerical plot for ($u,\alpha_{u}$) when using
(\ref{Heq})-(\ref{seq}). These flows show that inflation is emergent
from the saddle point ($\bullet$) (0,0) and is realized along the
red line. Finally, the inflation ends at the point
({\footnotesize$\blacksquare$}) ($-0.69,6$), which corresponds to
$\epsilon_{N}^{V}=1$.}
\end{figure}

Now, we solve  (\ref{Heq})-(\ref{seq}) and
(\ref{delu})-(\ref{delalphau}) numerically by considering  three  of
exponential, quadric (chaotic), and  quartic potentials.
\\
\\
$(i)~V=V_0e^{\lambda\phi}$\\
\\
For an exponential potential, the inflaton velocity (\ref{dotphiu})
is given by
\begin{eqnarray}\label{pdp1}
\dot{\phi}\simeq-\frac{M_{\rm P}^3\tilde{M}^2\lambda}{\sqrt{3V_0}}e^{-\frac{\lambda}{2}\phi},
\end{eqnarray}
which implies that when solving (\ref{Heq})-(\ref{seq}) numerically
for various initial conditions, any solution  should attract the
trajectory of (\ref{pdp1}) [see Fig.4 (left)]. Here we note that
there is no spiral sink because the potential approaches zero in the
limit of $\phi \to -\infty$.  We point out that $\alpha_u=-6u$
(\ref{ua}) cannot be an attractor (point) because of evolution of
$H$ in $\alpha_u$ (\ref{dimless1}) even though $V'/V=\lambda$. This
contrasts to the CC case, which yields an attractor (point) at
$x=-\alpha_V/6$ for the exponential potential [see Fig.1 (right)]
implying that the slow-roll never ends.

On the other hand, for the NDC case with the exponential potential,
one  checks that the inflationary attractor in the ($u,\alpha_u$) is
given by a slow-roll line of $\alpha_{u}=-6u$ [see Fig.4 (right)].
In this case, inflation can be realized as an emergence from the
saddle point $(0,0)$. The stream flows in the right  of Fig. 4 can
leave the  point $(0,0)$ only along the red line of $u=-\alpha_u/6$
corresponding to the slow-roll line  (\ref{ua}) and the inflation
ends at the point ($-0.69, 6$), which corresponds to
$\epsilon_{N}^{V}=1$ with $\alpha_u=6$.
\\
\\
$(ii)~V=V_0\phi^2$\\
\\
In case of chaotic potential, the inflaton velocity  (\ref{dotphiu})
is given by
\begin{eqnarray}\label{solph1}
\dot{\phi}\simeq\left\{\begin{array}{ll}
-\frac{2M_{\rm P}^3\tilde{M}^2}{\sqrt{3V_0}}\frac{1}{\phi^2},
~~(\phi>0) \label{soo1}\\
+\frac{2M_{\rm P}^3\tilde{M}^2}{\sqrt{3V_0}}\frac{1}{\phi^2},~~(\phi<0) \label{soo2}\end{array}\right..
\end{eqnarray}
Figure 5 shows that when solving  Eqs. (\ref{Heq})-(\ref{seq}) and
(\ref{delu})-(\ref{delalphau}) numerically, the inflationary
attractors are given by the red curve (\ref{solph1}) followed by
stable limited cycle on ($\phi,\dot{\phi}$) [Fig.5 (left)] and  the
slow-roll line (\ref{ua}) starting at (0,0) [Fig.5 (right)]. We note
here that the stable limited cycle where nearby curves spiral
towards closed curve $C$ appears instead of the spinal sink in the
CC. Hence, the trajectories trace out a closed curve $C$.

\begin{figure}[t!]
\begin{center}
\begin{tabular}{c}
\includegraphics[width=.85\linewidth,origin=tl]{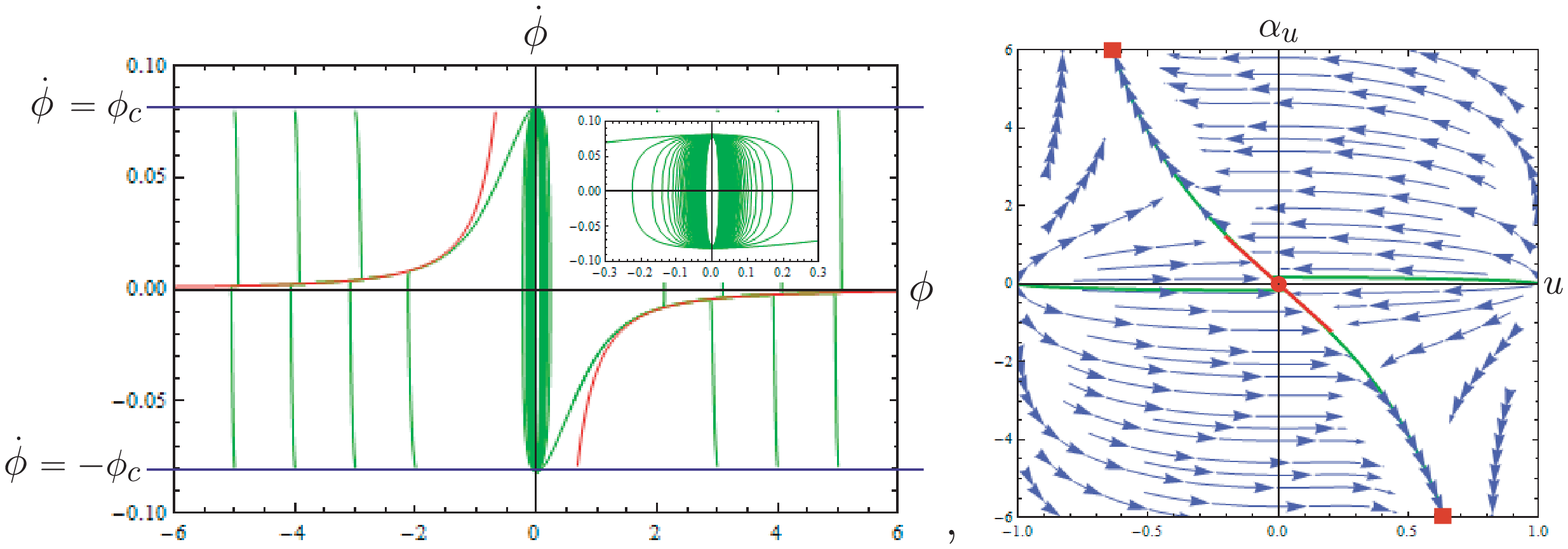}
\end{tabular}
\end{center}
\caption{The phase portrait $(\phi,\dot\phi)$ [left] and stream flow
$(u,\alpha_u)$ [right] for  $V=V_0\phi^2$ with $V_0=0.1$, $M_{\rm
P}=1$, $\tilde{M}=0.1$, and $\phi_c=M_{\rm
P}\tilde{M}\sqrt{2}/{\sqrt{3}}\simeq0.08$. Left panel shows the
trajectories toward the inflationary attractor (red curves) followed
by stable limited cycle. These curves correspond to a red line of
$\alpha_{u}=-6u$ in the right panel. Also, the green flows in the
right show  the numerical plot for ($u,\alpha_{u}$) when solving
(\ref{Heq})-(\ref{seq}). These flows indicate that the inflation is
emergent from the saddle point $(\bullet)$ (0,0) and is realized
along the red line. Finally, the inflation ends at the point
({\footnotesize$\blacksquare$}) ($\mp0.63,\pm6$), which corresponds
to $\epsilon_{N}^{V}=1$.}
\end{figure}

Inflation can be realized as an emergence from the saddle point
$(0,0)$. The stream flows in the right of Fig. 5 indicate that the
inflation  can leave the saddle point $(0,0)$ only along the red
line of $u=-\alpha_u/6$ corresponding to the slow-roll line
(\ref{ua}). The inflation ends at the point ($\mp0.63, \pm6$), which
corresponds to $\epsilon_{N}^{V}=1$ with $\alpha_u=\pm6$.
Furthermore, comparing the red curve in Fig.5 (left) with the red
line in Fig.2 (left) indicates that the NDC suppresses  the kinetic
term $\dot{\phi}$ whereas it enhances higher friction.
\\
\\
$(iii)~V=V_0\phi^4$\\
\\
The inflaton velocity  (\ref{dotphiu}) for this case is given by
\begin{eqnarray}\label{solphd3}
\dot{\phi}\simeq -\frac{4M_{\rm
P}^3\tilde{M}^2}{\sqrt{3V_0}}\frac{1}{\phi^3}.
\end{eqnarray}
Finally, Figure 6 shows that when solving (\ref{Heq})-(\ref{seq})
and (\ref{delu})-(\ref{delalphau}) numerically, the inflationary
attractors are given by the red curve (\ref{solphd3}) followed by
stable limited cycle on ($\phi,\dot{\phi}$) [Fig.6 (left)] and
slow-roll line (\ref{ua}) on ($u,\alpha_{u}$) [Fig.6 (right)].
Inflation can be realized as an emergence from the saddle  point
$(0,0)$. The stream flows in the right of Fig. 6 indicate that
inflation  can leave the point $(0,0)$ only along the red line of
$\alpha_u=-6u$ corresponding to the slow-roll line (\ref{ua}) and
the inflation ends at the point ($\mp0.66, \pm6$), which corresponds
to $\epsilon_{N}^{V}=1$ with $\alpha_u=\pm6$. Also, comparing the
red curve in Fig.6 (left) with the red curve in Fig.3 (left)
indicates that the NDC suppresses  the inflaton velocity
$\dot{\phi}$ whereas it enhances higher friction. This means that
the initial kinetic energy in the NDC  is small when compared with
the potential energy. On the other hand, the kinetic energy in the
CC rapidly decreases until eventually the potential energy dominates
and the universe may enter the slow-roll inflation. However, it is
not easy to obtain an enough e-folds number from $\phi^4$ in the CC.
This explains why this potential was ruled out~\cite{Ade:2015lrj}.
\begin{figure}[t!]
\begin{center}
\begin{tabular}{c}
\includegraphics[width=.85\linewidth,origin=tl]{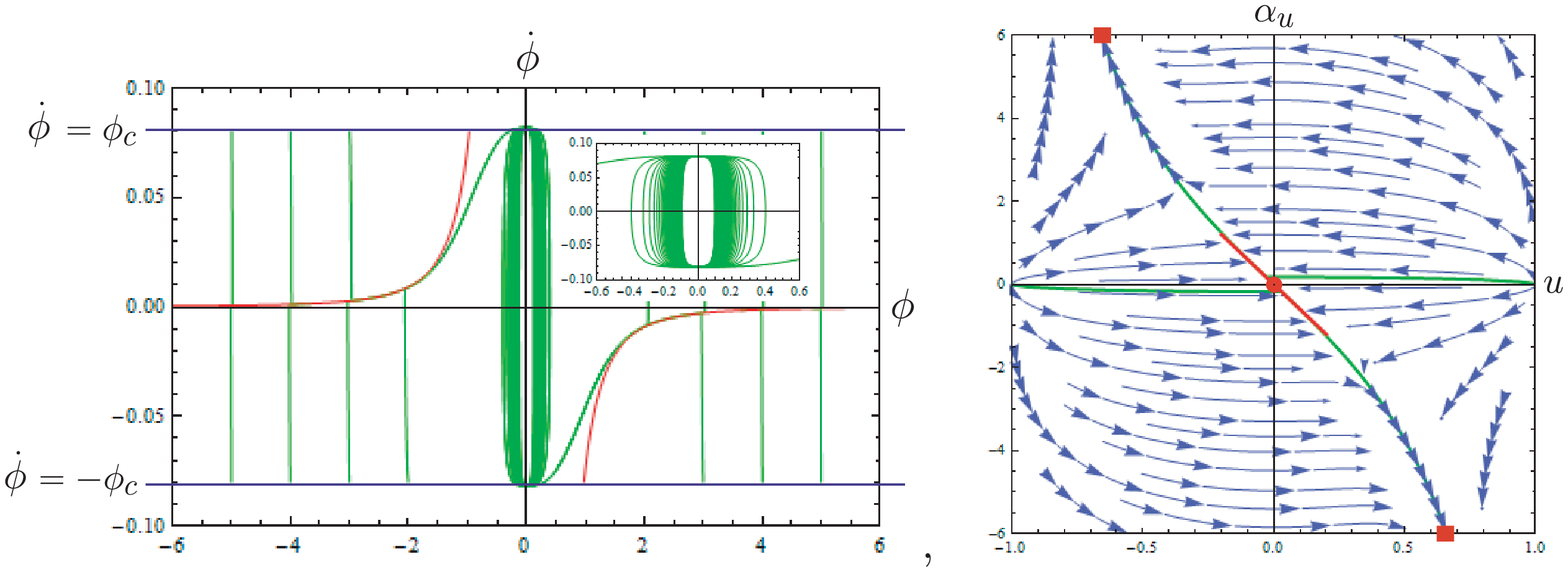}
\end{tabular}
\end{center}
\caption{The phase portrait $(\phi,\dot\phi)$ [left] and stream flow
$(u,\alpha_u)$ [right] for $V=V_0\phi^4$ with $V_0=0.1$, $M_{\rm
P}=1$, $\tilde{M}=0.1$, and $\phi_c=M_{\rm
P}\tilde{M}\sqrt{2}/{\sqrt{3}}\simeq0.08$. Left panel shows the
trajectories toward the inflationary attractor (red curves) followed
by stable limited cycle. These curves correspond to a red line of
$\alpha_{u}=-6u$ in the right panel. Also, the green flows in the
right  show the numerical plot for ($u,\alpha_{u}$) when using
(\ref{Heq})-(\ref{seq}). This flow shows that the inflation is
emergent from the saddle point $(\bullet)$ (0,0) and is realized
along the red line. Finally, the inflation ends at the point
({\footnotesize$\blacksquare$}) ($\mp0.66, \pm6$), which corresponds
to $\epsilon_{N}^{V}=1$.}
\end{figure}
\section{Summary and Discussions}
First of all,  we compare the NDC with the CC with the same
potentials. In the phase portrait ($\phi,\dot{\phi}$), the slow-roll
inflation is identified with the presence of inflationary attractor
(slow-roll equation) followed by spiral sink (0,0) (the ending of
inflation) for the CC and stable limited cycle centered at (0,0) for
the NDC. See Fig. 7 for  the different  behaviors of $\phi$ and
$\dot{\phi}$ between the CC and NDC during the evolution of the
universe.  On the other hand, the corresponding autonomous forms for
${\bf X}$=($x(u),\alpha_{V(u)}$) indicated that the slow-roll
inflation is realized by emerging a saddle point (0,0) (the
beginning of inflation) and inflation leaves this point along the
slow-roll equation. This is so because of $(x,\alpha_V)\simeq
(\dot{\ln [\phi]},\frac{1}{\phi})$ for $\phi^2$ and
$(\frac{\dot{\phi}}{\phi^2},\frac{1}{\phi^2})$ for $\phi^4$ in the
CC case. In the NDC case, $(u,\alpha_u)\simeq
(\dot{\phi},\frac{1}{\phi^2})$ and $(\dot{\phi},\frac{1}{\phi^3})$.
This implies that even though the limit of $\phi\to 0$ is not
implemented  by the autonomous system, the other limit of $\phi\to
\infty$ can be easily realized at the origin (0,0).
\begin{figure}[t!]
\begin{center}
\begin{tabular}{c}
\includegraphics[width=.90\linewidth,origin=tl]{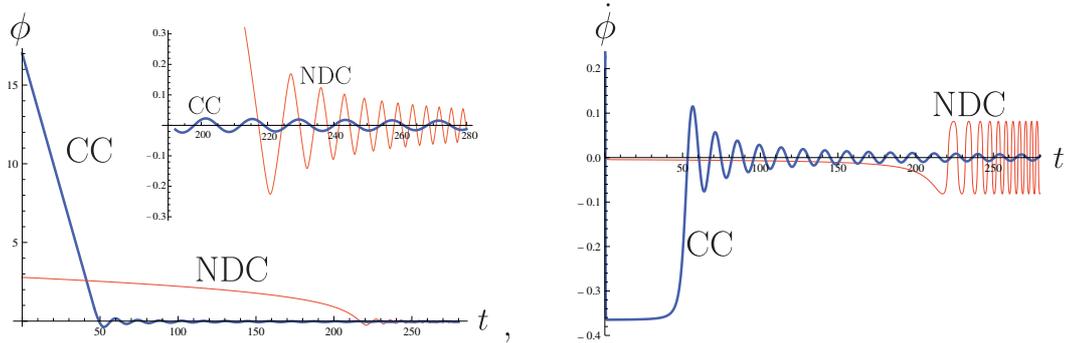}
\end{tabular}
\end{center}
\caption{The evolution of $\phi(t)$ [left] and $\dot{\phi}(t)$
[right] with respect to time $t$ for potential $\phi^2$. The left
figure shows that the inflaton varies little during large
inflationary period for the NDC, while it varies quickly during
small inflationary period for the CC. After inflation, $\phi$ decays
with oscillation for CC, while it oscillates rapidly for NDC. The
right one indicates that for large $t$, $\dot{\phi}$ oscillates
without damping for the NDC, while it oscillates with damping for
the CC.}
\end{figure}

 We
summarize our results in Table 1. $\phi_i$ is the value of $\phi$
when the phase space trajectory hits the inflationary  attractor
while $\phi_f$ is determined by $\epsilon_{C,N}^V=1$. The relations
are given for the exponential ($e^{\lambda\phi}$) and power-law
potential ($\phi^n$) by
\begin{eqnarray}
\phi_f-\phi_i=-\lambda M_{\rm P}^2 N~~~({\rm CC}),
&&e^{\lambda\phi_f}-e^{\lambda\phi_i}=-\lambda^2M_{\rm P}^4
\tilde{M}^2
\frac{N}{V_0}~~~({\rm NDC})\nonumber\\
\phi_f^2-\phi_i^2=-2M_{\rm P}^2 n N~~~({\rm CC}),
&&\phi_f^{n+2}-\phi_i^{n+2}=-n(n+2)M_{\rm P}^4 \tilde{M}^2
\frac{N}{V_0}~~~({\rm NDC})\nonumber,
\end{eqnarray}
where the e-fold number is $N=70$. $\Delta \phi=\phi_f-\phi_i$ in
the CC and NDC are  15.38 and 2.07 for $\phi^2$, 21.01 and 1.39 for
$\phi^4$, and not available (N.A.) and 4.94 for $e^{\lambda \phi}$.
During inflation, the inflaton must be in a slow-roll regime. This
implies that the inflaton varies little during the inflationary
phase and thus, it satisfies  $\dot{\phi}^2 \ll V$ and $\ddot{\phi}
\ll 3H\dot{\phi}$. In this sense, the NDC with potentials is easier
to make slow-roll inflation than the CC  with the same potentials.

Especially, the exponential (non-flat) potential is regarded as a
testing potential because it  provides a power-law inflation in the
CC model while it could provide a slow-roll inflation in the NDC. We
note that as was shown in Fig.1 (right), there is no saddle point as
a starting point of slow-roll line even though (0,1) denotes a
saddle point. Also, there is no stable limited cycle  in the phase
portrait [Fig.4 (left)] of the NDC which features the exponential
potential.

\begin{table*}[t]
\begin{center}
\begin{tabular}{|c|c|c|c|c|c|c|}
 \hline
  & \multicolumn{4}{c|}{CC} \\
\hline
potential  & $\phi_i$ & $\phi_f$ & spiral sink  & saddle point \\
      \hline
$\phi^2$ & $16.79$ & 1.41 &  O & O \\
\hline
$\phi^4$  &23.83 & 2.82 & O &O \\
\hline
$e^{\lambda\phi}$ & N.A. & N.A. &  X& X \\
\hline\hline
& \multicolumn{4}{c|}{NDC} \\
\hline
potential  & $\phi_i$ & $\phi_f$  & stable limited cycle &  saddle point  \\
      \hline
$\phi^2$ & 2.74 &0.67 & O&O \\
\hline
$\phi^4$  & 2.35 & 0.96 & O&O \\
\hline
$e^{\lambda\phi}$ & 1.95 & -2.99&X &O\\
\hline
\end{tabular}
\end{center}
\caption[crit]{Summary on how inflation takes place for the CC and
NDC with three potentials of $\phi^2,\phi^4,$ and $e^{\lambda
\phi}$. $\phi_i$ denotes minimal value satisfying the e-folds
$N\ge70$ and $\phi_f$ is the end of inflation. The presence of
either spinal sink or stable limited cycle  represents the end of
inflation in the phase portrait ($\phi,\dot{\phi}$), while the
saddle point denotes the emergence of inflation in the autonomous
form.}
\end{table*}

In order to see how slow-roll inflation occurs in the NDC case, we
compare the NDC with the CC when one chooses the chaotic potential
$V=V_0\phi^2$.  Figure 7 (left) indicates that the inflaton varies
little during large slow-roll period for the NDC, while it varies
quickly during small slow-roll period for the CC. This implies that
the NDC would inflate enough comparing with the CC. After the
inflation [see Fig.7 (right)], the velocity of the inflation
$\dot{\phi}$ oscillates with respect time in the NDC, while its
velocity describes a damped-oscillation in the CC. This
distinguishes  the stable limited cycle for the NDC from the spinal
sink for the CC.

It is worth noting that  there are three phases
($\longrightarrow\bullet\longrightarrow$spiral sink) in the CC
case~\cite{Donoghue:2007ze}: i) Initially, kinetic energy dominates
[see Fig.2 (left) and Fig.3 (left)]. ii) Due to the rapid decrease
of the kinetic energy, the trajectory runs  to the inflationary
attractor line (\ref{solph11}). All initial trajectories are
attracted to this line, which is the key feature of slow-roll
inflation. iii) At the end of inflation, the magnitude of inflaton
velocity $|\dot{\phi}|$ decreases. There is inflaton decay and
reheating which shows the appearance of spiral sink. On the other
hand, three stages ($\longrightarrow\bullet\longrightarrow$stable
limited cycle) in the NDC are as follows: i) Initially, potential
energy dominates [see Fig.5 (left) and Fig.6 (left)]. ii) Due to the
gravitationally enhanced friction, all initial trajectories  are
attracted quickly to the inflationary attractor. iii) At the end of
inflation, the magnitude of the inflation velocity $|\dot{\phi}|$
increases. Then, the stable limited cycle appears, which differs
from the spiral sink in the CC case. This is a rapidly oscillating
phase \cite{Sadjadi:2013psa}, but it appears after slow-roll
inflation \cite{Jinno:2013fka}.

We would like here to warn the reader that, because of numerical
complexity, all our analysis are either made in the NDC or in the CC
system but not in the more physical combined NDC+CC system (although
some hints are given in the Appendix). Nevertheless, we expect the
sector we have excluded to provide an even larger space of initial
conditions leading to inflation. Thus, our results although limited
already brings the important message that the NDC coupling
facilitate inflation with respect to the sole CC one.

Finally, we wish to mention that  for the power-law potential
$\phi^n$, the field excursion of the inflaton is sub-Planckian due
to the NDC~\cite{Yang:2015pga}. Thus, the tensor-to-scalar ratio $r$
is a factor of $(n+2)/2$ smaller than the results in the CC which
brings the quartic and quadric potentials to be consistent with the
observation at the 95$\%$ CL.

\vspace{0.35cm}

 {\bf Acknowledgement}

\vspace{0.25cm}
 We thank Wonwoo Lee and Seoktae Koh for useful discussions. Y.Myung and T.Moon were supported by the National
Research Foundation of Korea (NRF) grant funded by the Korea
government (MEST) (No.2012-R1A1A2A10040499). B.Lee was supported by
 the National Research Foundation of Korea (NRF) grant funded by
the Korea government (MSIP) (2014R1A2A1A01002306).

\section*{Appendix}
\appendix
\section{The autonomous system for CC+NDC}

In this appendix, we consider an action including a canonical
coupling term as
\begin{eqnarray}
S_{\rm}=\int d^4x \sqrt{-g}\Big[\frac{M_{\rm
P}^2}{2}R-\frac{1}{2}\left(g_{\mu\nu}-\frac{1}{\tilde{M}^2}G_{\mu\nu}\right)
\partial^{\mu}\phi\partial^{\nu}\phi-V(\phi)\Big].
\end{eqnarray}
For a spatially flat spacetime (\ref{deds1}) with $\phi=\phi(t)$,
the Friedmann equations and scalar equation are given by
\begin{eqnarray}
H^2&=&\frac{1}{3M_{\rm P}^2}\Big[\frac{1}{2}\dot{\phi}^2\Big(1+\frac{9H^2}{\tilde{M}^2}\Big)+V\Big],\label{aHeq}\\
&&\nonumber\\
\dot{H}&=&-\frac{1}{2M_{\rm
P}^2}\Big[\dot{\phi}^2\Big(1+\frac{3H^2}{\tilde{M}^2}-\frac{\dot{H}}{\tilde{M}^2}\Big)-\frac{2H}{\tilde{M}^2}\dot{\phi}\ddot{\phi}
\Big],\label{adHeq}\\
&&\nonumber\\
&&\hspace*{-5em}\Big(1+\frac{3H^2}{\tilde{M}^2}\Big)\ddot{\phi}+3H\Big(1+\frac{3H^2}{\tilde{M}^2}+\frac{2\dot{H}}{\tilde{M}^2}\Big)\dot{\phi}+V'=0\label{aseq}.
\end{eqnarray}
To perform the dynamical analysis for the equations
(\ref{aHeq})-(\ref{aseq}), we first introduce the following
dimensionless quantities,
\begin{eqnarray}
x\equiv\frac{\dot{\phi}}{M_{\rm P}\sqrt{6} H},
~u\equiv\sqrt{\frac{3}{2}}\frac{\dot{\phi}}{M_{\rm
P}\tilde{M}},~y\equiv\frac{\sqrt{V}}{M_{\rm P}\sqrt{3}
H},~\alpha_{V}=\sqrt{6}M_{\rm
P}\frac{V'}{V},~\alpha_{u}=\sqrt{6}\frac{M_{\rm
P}\tilde{M}V'}{VH}.\label{adimless}
\end{eqnarray}
Here, $x$ and $u$ represent the dominance of the kinetic term for CC
and the kinetic term for NDC, while $y$  denotes the dominance of
the potential term for CC and NDC. However, a constraint equation
obtained from  Eq. (\ref{aHeq}) is given by
\begin{eqnarray}
 x^2+ u^2+y^2=1,
\end{eqnarray}
which allows us to eliminate $y$.

Introducing $N=\ln{a}$, (\ref{aHeq})-(\ref{aseq}) can be written in
terms of the quantities (\ref{adimless}) being consisting the
following autonomous form for ${\bf X}=(x,u,\alpha_V,\alpha_u)$:
\begin{eqnarray}
\frac{dx}{dN}&=&x(\epsilon+\delta),\label{adelx}\\
\frac{du}{dN}&=&u\delta,\label{adelu}\\
\frac{d\alpha_{V}}{dN}&=&\alpha_{V}^2x(\Gamma-1),\label{adelalphav}\\
\frac{d\alpha_{u}}{dN}&=&\epsilon\alpha_{u}+\frac{1}{3}(\Gamma-1)u\alpha_{u}^2,\label{adelalphau}
\end{eqnarray}
where $\epsilon$, $\delta$, and their relation are given by
\begin{eqnarray}
\epsilon=-\frac{\dot{H}}{H^2},~\delta=\frac{\ddot{\phi}}{\dot{\phi}H},~\epsilon=\frac{3x^2+u^2-\frac{2}{3}u^2\delta}{1-\frac{1}{3}u^2}.
\end{eqnarray}
It is not an easy  task  to analyze the autonomous system for ${\bf
X}=(x,u,\alpha_V,\alpha_u)$ (CC+NDC) completely because of its
complexity.   However, we expect to extract some information on  the
full system (CC+NDC) by analyzing   fixed points  obtained from  the
autonomous system (\ref{adelx})-(\ref{adelalphau}) whose compact
form is rewritten  by ${\bf X}'={\bf f}({\bf X})$.  The fixed points
${\bf X}^f$ are extracted from the condition of ${\bf X}'=0$ and
they provide qualitative information on the global dynamics of the
system, independently of the initial conditions and specific
evolution of the system. This information might include all previous
fixed points  and new fixed points  from other combinations. First
of all, we have recovered all previous fixed points  from ${\bf
X}'=0$ and summarize them (P1$_\pm\sim$P8) in Table 2. Then, what
are new fixed points which might be found from other combinations?
We expect to have three candidates from the analysis of ${\bf
X}'=0$.  It is checked easily that the two kinetic-dominant fixed
point of ($x^f$,0,$u^f$,0,$y^f\approx0$) is not allowed for the
autonomous system (\ref{adelx})-(\ref{adelalphau})  because both CC
and NDC are kinetic term. The two remaining points are the
potential-dominant fixed points of
(0,$\alpha_V^f$,0,$\alpha_u^f$,$y^f$) and
(0,$\alpha_V^f$,$u^f$,$\alpha_u^f$,$y^f$). Explicitly, the former
includes two saddle points P9 and P10,  while the latter is given by
 repeller  P11$_\pm$ which is similar to P4$_\pm$. Finally, it is worth to  note that  P9
 and P10  might  represent slow-roll inflation in the full system (CC+NDC) because these belong to
 saddle points.

\begin{table*}[t]
\begin{center}
\begin{tabular}{|c|c|c|c|c|c|c|c|}
 \hline
  fixed points  & $(x,\alpha_V,u,\alpha_u,y)$ & potential  & type  & stability &where\\
\hline
 P1$_\pm$  & ($\pm1$,$\sqrt{6}$,0,0,0) & $V_0 e^{\phi}$ & CC  & R &Fig.1(right)\\
      \hline
P2 & ($-1/\sqrt{6}$,$\sqrt{6}$,0,0,$\sqrt{5/6}$) & $V_0 e^{\phi}$  & CC & A &Fig.1(right)\\
\hline
P3  &(0,$\sqrt{6}$,0,0,1) & $V_0 e^{\phi}$  & CC & S &Fig.1(right)\\
\hline
P$4_\pm$  &(0,$\sqrt{6}$,$\pm1$,0,0) & $V_0 e^{\phi}$  & NDC & R &Fig.4(right)\\
\hline
P5  &(0,$\sqrt{6}$,0,0,1) & $V_0 e^{\phi}$  & NDC & S &Fig.4(right)\\
\hline
P6  & (0,0,0,0,1) & $V_0 \phi^p$ & CC  & S &Fig.2,3(right)\\
\hline
P7$_\pm$  & (0,0,$\pm1$,0,0) & $V_0 \phi^p$ & NDC  & R &Fig.5,6(right)\\
\hline
P8  & (0,0,0,0,1) & $V_0 \phi^p$ & NDC  & S &Fig.5,6(right)\\
\hline
P9 & (0,$\sqrt{6}$,0,$\alpha_u$,1) & $V_0 e^{\phi}$ & CC+NDC & S & new\\
\hline
P10 & (0,$\alpha_V$,0,$\alpha_u$,1) & $V_0 \phi^p$ & CC+NDC & S & new\\
\hline
P11$_\pm$ & (0,$\alpha_V$,$\pm1$,0,0) & $V_0 \phi^p$ & CC+NDC & R & new\\
\hline
\end{tabular}
\end{center}
\caption[crit]{List of all previous and new fixed points ${\bf X}^f$
in the autonomous system of CC+NDC for $V=V_0e^\phi$ and
$V=V_0\phi^p(p=2,4)$. In the stability, A, S, and R represent
attractor, saddle point, and repeller, respectively where R means
kinetic-dominant fixed point ($y=0$) and S implies
potential-dominant fixed point ($y=1$). }
\end{table*}


\newpage



\begin{thebibliography}{99}
\bibitem{Amendola:1993uh}
  L.~Amendola,
   Phys.\ Lett.\ B {\bf 301}, 175 (1993)  [gr-qc/9302010].

\bibitem{Sushkov:2009hk}
  S.~V.~Sushkov,
  Phys.\ Rev.\ D {\bf 80}, 103505 (2009)  [arXiv:0910.0980 [gr-qc]].



\bibitem{Germani:2010gm}
  C.~Germani and A.~Kehagias,
   Phys.\ Rev.\ Lett.\  {\bf 105}, 011302 (2010)  [arXiv:1003.2635 [hep-ph]].

\bibitem{Germani:2011ua}
  C.~Germani and Y.~Watanabe,
  JCAP {\bf 1107}, 031 (2011)  [Addendum-ibid.\  {\bf 1107}, A01 (2011)]  [arXiv:1106.0502 [astro-ph.CO]].





\bibitem{Germani:2011mx}
  C.~Germani,
  Rom.\ J.\ Phys.\  {\bf 57}, 841 (2012)  [arXiv:1112.1083 [astro-ph.CO]].

\bibitem{PU} P. Peter and J.-P. Uzan, The Primordial Cosmology
              (Oxford Univ., Oxford, 2009)  p. 463.






\bibitem{Dalianis:2014nwa}
  I.~Dalianis and F.~Farakos,
   Phys.\ Rev.\ D {\bf 90}, no. 8, 083512 (2014)  [arXiv:1405.7684 [hep-th]].

\bibitem{Copeland:1997et}
  E.~J.~Copeland, A.~R.~Liddle and D.~Wands,
  Phys.\ Rev.\ D {\bf 57}, 4686 (1998)  [gr-qc/9711068].

\bibitem{Leon:2014rra}
  G.~Leon and C.~R.~Fadragas,
  arXiv:1412.5701 [gr-qc].  


\bibitem{Leon:2008de}
  G.~Leon,
  Class.\ Quant.\ Grav.\  {\bf 26}, 035008 (2009)  [arXiv:0812.1013 [gr-qc]].

\bibitem{Fadragas:2014mra}
  C.~R.~Fadragas and G.~Leon,
  Class.\ Quant.\ Grav.\  {\bf 31}, no. 19, 195011 (2014)  [arXiv:1405.2465 [gr-qc]].

\bibitem{Cai:2014uka}
  Y.~F.~Cai, J.~O.~Gong, S.~Pi, E.~N.~Saridakis and S.~Y.~Wu,
   arXiv:1412.7241 [hep-th].


\bibitem{Felder:2002jk}
  G.~N.~Felder, A.~V.~Frolov, L.~Kofman and A.~D.~Linde,
  Phys.\ Rev.\ D {\bf 66}, 023507 (2002)  [hep-th/0202017].





\bibitem{Kofman:2002cj}
  L.~Kofman, A.~D.~Linde and V.~F.~Mukhanov,
  JHEP {\bf 0210}, 057 (2002)  [hep-th/0206088].





\bibitem{Hindmarsh:2011hx}
  M.~Hindmarsh, D.~Litim and C.~Rahmede,
  JCAP {\bf 1107}, 019 (2011)  [arXiv:1101.5401 [gr-qc]].

\bibitem{Contillo:2011ag}
  A.~Contillo, M.~Hindmarsh and C.~Rahmede,
  Phys.\ Rev.\ D {\bf 85}, 043501 (2012)  [arXiv:1108.0422 [gr-qc]].


\bibitem{Yang:2015pga}
  N.~Yang, Q.~Gao and Y.~Gong,
  arXiv:1504.05839 [gr-qc].



\bibitem{Ade:2015lrj}
  P.~A.~R.~Ade {\it et al.}  [Planck Collaboration],
  arXiv:1502.02114 [astro-ph.CO].


\bibitem{Tsujikawa:2012mk}
  S.~Tsujikawa,
   Phys.\ Rev.\ D {\bf 85}, 083518 (2012)  [arXiv:1201.5926 [astro-ph.CO]].




\bibitem{Feng:2014tka}
  K.~Feng and T.~Qiu,
  Phys.\ Rev.\ D {\bf 90}, no. 12, 123508 (2014)  [arXiv:1409.2949 [hep-th]].




\bibitem{Skugoreva:2013ooa}
  M.~A.~Skugoreva, S.~V.~Sushkov and A.~V.~Toporensky,
  Phys.\ Rev.\ D {\bf 88}, 083539 (2013)  [Phys.\ Rev.\ D {\bf 88}, no. 10, 109906 (2013)]  [arXiv:1306.5090 [gr-qc]].

\bibitem{Germani:2011bc}
  C.~Germani, L.~Martucci and P.~Moyassari,
  Phys.\ Rev.\ D {\bf 85}, 103501 (2012)
  [arXiv:1108.1406 [hep-th]].



\bibitem{Germani:2010ux}
  C.~Germani and A.~Kehagias,
  JCAP {\bf 1005} (2010) 019
   [JCAP {\bf 1006} (2010) E01]
  [arXiv:1003.4285 [astro-ph.CO]].



\bibitem{Donoghue:2007ze}
  J.~F.~Donoghue, K.~Dutta and A.~Ross,
  Phys.\ Rev.\ D {\bf 80}, 023526 (2009)  [astro-ph/0703455 [ASTRO-PH]].




\bibitem{Sadjadi:2013psa}
  H.~M.~Sadjadi and P.~Goodarzi,
  Phys.\ Lett.\ B {\bf 732} (2014) 278
  [arXiv:1309.2932 [astro-ph.CO]].


\bibitem{Jinno:2013fka}
  R.~Jinno, K.~Mukaida and K.~Nakayama,
  JCAP {\bf 1401} (2014) 01,  031
  [arXiv:1309.6756 [astro-ph.CO]].


\end{thebibliography}
\end{document}